\documentclass[12pt]{article}
\usepackage[plainpages=false,pdfpagelabels,unicode]{hyperref}
\usepackage[english]{babel}
\usepackage{comment}
\usepackage{amsfonts}
\usepackage{graphicx}
\usepackage{epstopdf}
\usepackage{epsfig}
\usepackage{amssymb}
\usepackage{setspace}
\usepackage{caption}
\usepackage{mathrsfs}
\usepackage{mathtools}
\usepackage{color}
\usepackage{amsmath}
\usepackage{amsthm}
\usepackage{float}
\usepackage[utf8]{inputenc}
\usepackage[square,comma,numbers,sort&compress]{natbib}

\newcommand\numberthis{\addtocounter{equation}{1}\tag{\theequation}}
\newcommand {\be}{\begin{equation}}
\newcommand {\ee}{\end{equation}}
\newcommand {\te}{\mathrm}
\newcommand {\bea}{\begin{array}}

\newcommand {\eea}{\end{array}}
\renewcommand {\theequation}{\thesection.\arabic{equation}}

\numberwithin{equation}{section}

\newcommand{\sch}{Schwarzschild }
\begin{document}
\begin{titlepage}
\vspace{1cm} 
\begin{center}
{\Large \bf {Equatorial Ba\~nados-Silk-West effect in Kerr-Newman-Taub-NUT spacetime revisited}}\\
\end{center}
\vspace{1cm}
\begin{center}
\renewcommand{\thefootnote}{\fnsymbol{footnote}}
Delvydo Melvernaldo{\footnote{delvydo@zoho.com or 7216007@student.unpar.ac.id}}\\
Center for Theoretical Physics, Department of Physics,\\
Parahyangan Catholic University, Bandung 40141, Indonesia\\
\renewcommand{\thefootnote}{\arabic{footnote}}
\end{center}

\begin{abstract}
In this paper, we revisit the possibilities of Ba\~nados-Silk-West (BSW) effect in Kerr-Newman-Taub-NUT (KNTN) spacetime for two neutral particles moving over the equatorial plane and constant $\theta$ in Boyer-Lindquist coordinate.
Contrary to a previous study on this topic, we found that BSW effect for two particles confined to move over the equatorial plane is not possible. Numerical calculations shows that BSW effect in constant $\theta$ geodesics is possible under certain circumstances.
\end{abstract}
\end{titlepage}
\onecolumn
\bigskip 

\section{Introduction}
\label{sec:intro}

Kerr-Newman-Taub-NUT spacetime is a solution to Einstein-Maxwell field equations that describes a rotating electrically charged mass with NUT parameter \cite{Griffiths_2009, demianski1966combined}. The spacetime contains four parameters that describes the gravitational mass, the rotation parameter, the electric charge, and the NUT parameter. The NUT parameter is commonly interpreted as the gravito-magnetic monopole moment which is an addition to the gravitoelectric monopole moment (mass) as a more generalized \sch solution \cite{Newman_1963, Bini_2003}, or as a "twist parameter" of the electromagnetic field \cite{Al_Badawi_2006}.

The authors of \cite{Ba_ados_2009} showed that a collision in a equatorial plane near the Kerr black hole can happen with an arbitrarily high center of mass energy. This effect is now called the BSW effect and was further studied in the Kerr spacetime in literature \cite{Berti_2009, Jacobson_2010, Galajinsky_2013, Zaslavskii_2010, Harada_2011, Harada_2011_ISCO, Harada_2014, Lake_2010, Grib_2011, Grib_2011_Near, Grib_2010}. The same effect was then studied in other spacetime metrics, \cite{Wei_2010, Liu_2013} for the Kerr-Newman spacetime, literature \cite{Chakraborty_2015, Frolov_2012, Liu_2011, Chakraborty_2014_Strong} studies the Kerr-Taub-NUT spacetime, then article \cite{Zakria_2015} studies the Kerr-Newman-Taub-NUT (KNTN) spacetime which is the spacetime that we are going to discuss in this paper. One of the conclusions in the article \cite{Zakria_2015} is that BSW effect in the equatorial plane is possible. However, once the NUT parameter is chosen, equatorial geodesics are not guaranteed to exist and this was not taken into account in \cite{Zakria_2015}.

Literature \cite{Bini_2003, Cebeci_2016, Cebeci_2019} stated that equatorial geodesics in KNTN spacetime can exist in certain cases, i.e. arbitrary values of rotation parameter and NUT parameter of the spacetime does not guarantee the existence of equatorial geodesics. The nature of equatorial geodesics in KNTN spacetime was explored further in \cite{Cebeci_2019}. From those studies, it is clear that we need to impose more constraints than previously applied in \cite{Zakria_2015} to analyze BSW effect for the equatorial plane case in KNTN spacetime completely. We will also try to analyze the potential of BSW effect in KNTN spacetime with constant $\theta$ geodesics.

The paper is organized as follows. In section \ref{sec:KNTN}, we will give an overview of the KNTN spacetime, equation of motions, and center-of-mass energy of two test particles in KNTN spacetime. In section \ref{sec:const-theta}, we will obtain the necessary constraints to ensure the existence of geodesics with constant $\theta$. In section \ref{sec:BSW-KNTN}, we will apply the constraints obtained from section \ref{sec:const-theta}, the possibility of BSW effect for two neutral test particles with equatorial geodesics and for two neutral test particles with any arbitrary constant $\theta$ geodesics. Throughout this paper, we use the units such that $c=G=1$.

\section{Geodesics in Kerr-Newman-Taub-NUT spacetime}
\label{sec:KNTN}

We study the Einstein-Maxwell system dictated by Einstein's field equations
\begin{equation}
	\label{eq:Eins-Field}
	G_{\mu\nu} = R_{\mu\nu} - \frac{1}{2} g_{\mu\nu} R = 8 \pi T_{\mu\nu}
\end{equation}
where $T_{\mu\nu}=\frac{1}{4\pi}\left( F_{\mu\alpha} F_\nu^{\ \alpha} - \frac{1}{4} g_{\mu\nu} F_{\alpha\beta} F^{\alpha\beta} \right)$ and the source-free condition $\nabla_\mu F^{\mu\nu}=0$.
Kerr-Newman-Taub-NUT spacetime is a solution to such system and can be written in Boyer-Lindquist coordinate $x^\mu = (t, r, \theta, \phi)$ as in \cite{Grenzebach_2014, Bini_2003, Miller_1973, Zakria_2015}
\begin{equation}
    \begin{split}
        \text{d}s^2 = {} & -\frac{1}{\Sigma} (\Delta - a^2 \sin^2\theta) \text{d}t^2
        + \frac{2}{\Sigma} \left(\chi \Delta - a(\Sigma + a \chi) \sin^2\theta \right) \text{d}t\ \text{d}\phi\\
        &+ \frac{1}{\Sigma} \left( (\Sigma + a \chi)^2 \sin^2\theta - \chi^2 \Delta \right) \text{d}\phi^2
        + \frac{\Sigma}{\Delta} \text{d}r^2 + \Sigma \text{d}\theta^2
    \end{split}
    \label{eq:KNTN-metric}
\end{equation}
with
\begin{equation}
    \begin{split}
        \Sigma &= r^2 + (n + a \cos \theta)^2, \\
        \Delta &= r^2 - 2Mr - n^2 + a^2 + Q^2, \\
        \chi &= a \sin^2\theta - 2n \cos \theta,
    \end{split}
    \label{eq:sigma-delta-chi}
\end{equation}
where $M$ is the gravitational mass, $a$ is the rotational parameter, $Q$ is the electric charge of the gravitating object, and $n$ is the NUT parameter. This solution, which originated from \cite{demianski1966combined} and a special case from the Carter class \cite{Carter_1968a}, has a vector potential field $A$ which can be written in differential form notation as
\begin{equation}
    A = - \frac{Qr}{\Sigma} \left( \text{d}t - \chi \text{d}\phi \right).
    \label{eq:A-1form}
\end{equation}

We can see that the metric is singular at $\Sigma = 0$ and $\Delta = 0$. However, as \cite{Zakria_2015} pointed out, the Kretschmann scalar $R_{\mu\nu\alpha\beta} R^{\mu\nu\alpha\beta}$ suggests that the singularity at $\Delta=0$ is a coordinate singularity while $\Sigma = 0$ is a curvature singularity of the spacetime.

For the rest of this section, we are going to follow the derivation in \cite{Zakria_2015} to get the equations motion. We can use the Lagrangian
\begin{equation}
    \mathscr{L} = \frac{1}{2} g_{\mu\nu} \Dot{x}^\mu \Dot{x}^\nu
    \label{eq:lagrangian}
\end{equation}
using an affine parameter $\lambda$ with a relation $\lambda = \frac{\tau}{m}$ and $\dot{x}^\mu = \frac{\text{d}x^\mu}{\text{d}\lambda}$ where $\tau$ is the proper time and $m$ is the particle's rest mass. For timelike geodesics, the following parametrization condition must be satisfied.
\begin{equation}
    \frac{1}{m^2} g_{\mu\nu} \Dot{x}^\mu \Dot{x}^\nu = u^\mu u_\mu = -1.
    \label{eq:4vel-magnitude}
\end{equation}
The 4-momentum expressed in terms of the Lagrangian and its relation with the 4-velocity $u_\mu$ is
\begin{equation}
    P_\mu = \frac{\partial \mathscr{L}}{\partial \Dot{x}^\mu} = g_{\mu\nu} \dot{x}^\mu,
    \label{eq:4momentum}
\end{equation}
\begin{equation}
    u_\mu = \frac{P_\mu}{m},
    \label{eq:4vel}
\end{equation}
where $u^\mu=\frac{\mathrm{d}x^\mu}{\mathrm{d}\tau}$. With these, the Hamiltonian is
\begin{equation}
    \mathcal{H} = P_\mu \Dot{x}^\mu - \mathscr{L} = \frac{1}{2} g_{\mu\nu} \Dot{x}^\mu \Dot{x}^\nu = \frac{1}{2} g^{\mu\nu} P_\mu P_\nu
    \label{eq:hamiltonian}
\end{equation}
with Hamilton equations
\begin{equation}
    \Dot{x}^\mu = \frac{\partial \mathcal{H}}{\partial P_\mu},
    \Dot{P}_\mu = -\frac{\partial \mathcal{H}}{\partial x^\mu},
    \label{eq:hamilton-eqs}
\end{equation}
and Hamilton-Jacobi equation
\begin{equation}
    \mathcal{H} = -\frac{\partial S}{\partial \lambda}
    = \frac{1}{2} g^{\mu\nu} \frac{\partial S}{\partial x^\mu}
    \frac{\partial S}{\partial x^\nu}
    \label{eq:hamilton-jacobi}
\end{equation}
with $S$ is the Jacobi action and
\begin{equation}
    \frac{\partial S}{\partial x^\mu} = P_\mu.
    \label{eq:jacobi-action-4mom}
\end{equation}
The variables in the Jacobi action $S$ can be separated by taking (\ref{eq:4vel-magnitude}) into consideration and the fact that the metric (\ref{eq:KNTN-metric}) is not a function of $t$ and $\phi$ into
\begin{equation}
    S(t,r,\theta,\phi) = \frac{1}{2} m^2 \lambda - E m t + L m \phi + S_r(r) + S_\theta(\theta)
    \label{eq:jacobi-action}
\end{equation}
where $E$ is the specific energy of the particle, $L$ is the specific angular momentum of the particle, $S_r$ is a function that only depends on $r$, and $S_\theta$ is a function that only depends on $\theta$.

From equation (\ref{eq:jacobi-action-4mom}) and equation (\ref{eq:jacobi-action}) for the $t$ component, we get
$$E = -\frac{1}{m} \frac{\partial S}{\partial t} = -\frac{P_t}{m} = -u_t = -g_{\mu t} u^\mu,$$
\begin{equation}
    E = \frac{\Delta - a^2 \sin^2\theta}{\Sigma}u^t
    - \frac{\chi \Delta - a (\Sigma + a\chi) \sin^2\theta}{\Sigma}u^\phi
    \label{eq:eom-energy-specific}
\end{equation}
and for the $\phi$ component, we get
$$L = \frac{1}{m} \frac{\partial S}{\partial \phi} = \frac{P_\phi}{m} = u_\phi = g_{\mu \phi} u^\mu,$$
\begin{equation}
    L = \frac{\chi \Delta - a(\Sigma+a\chi)\sin^2\theta}{\Sigma}u^t
    + \frac{(\Sigma + a\chi)^2 \sin^2\theta - \chi^2 \Delta}{\Sigma}u^\phi.
    \label{eq:eom-angular-mom-specific}
\end{equation}
Solving (\ref{eq:eom-energy-specific}) and (\ref{eq:eom-angular-mom-specific}), we get the $t$ and $\phi$ components of the 4-velocity
\begin{equation}
    u^t = \frac{\chi (L-\chi E)}{\Sigma \sin^2\theta}
    + \frac{(\Sigma + a\chi) \left[ E(\Sigma+a\chi)\sin^2\theta \right] }{\Delta \Sigma}
    \label{eq:4vel-t}
\end{equation}
and
\begin{equation}
    u^{\phi} = \frac{L-\chi E}{\Sigma \sin^2\theta}
    + \frac{a \left[ (\Sigma + a\chi)E - aL \right] }{\Delta \Sigma}.
    \label{eq:4vel-phi}
\end{equation}
To get the $r$ and $\theta$ components of the 4-velocity, we use (\ref{eq:hamilton-jacobi}) and (\ref{eq:jacobi-action})
\begin{equation}
    -\frac{m^2}{2} = \frac{1}{2} \left[ m^2 g^{tt} E^2 +m^2 g^{\phi\phi} L^2 - 2m^2 g^{t\phi} E L + g^{rr} \left(\frac{\partial S_r}{\partial r}\right)^2 + g^{\theta\theta} \left(\frac{\partial S_\theta}{\partial \theta}\right)^2 \right].
    \label{eq:hamilton-jacobi-applied}
\end{equation}
With $\Sigma + a\chi = r^2+n^2+a^2$, we can separate functions of $\theta$ and functions of $r$ and get a constant which is called the Carter constant denoted with $K$ \cite{Carter_1968b}
\begin{gather*}
    \frac{1}{m^2} \left(\frac{\partial S_\theta}{\partial \theta}\right)^2
    + \cos^2\theta \left((1-E^2)a^2 + \frac{L^2}{\sin^2\theta}\right)
    + 2an \cos\theta (1-2E^2) + \frac{4n\cos\theta E}{\sin^2\theta}
    \left(n \cos\theta E + L\right)\\
    = -\frac{\Delta}{m^2} \left(\frac{\partial S_r}{\partial r}\right)^2
    - r^2 - n^2 - (L-aE)^2 + \frac{1}{\Delta}
    \left( (r^2+n^2+a^2)E - aL \right)^2 = K. \numberthis
    \label{eq:temp}
\end{gather*}
Using $u_r = \frac{1}{m} \frac{\partial S_r}{\partial r}$ and $u_\theta = \frac{1}{m} \frac{\partial S_\theta}{\partial \theta}$, we can get
\begin{equation}
    \Sigma u^{\theta} = \pm \sqrt{\Theta},
    \label{eq:4vel-theta}
\end{equation}
\begin{equation}
    \Sigma u^{r} = \pm \sqrt{R},
    \label{eq:4vel-r}
\end{equation}
with
\begin{equation}
    \begin{split}
        \Theta = \Theta(\theta) =
        {} & K - \cos^2\theta \left( (1-E^2)a^2 + \frac{L^2}{\sin^2\theta} \right) - 2an \cos \theta (1-2E^2) \\
        & - \frac{4n \cos \theta E}{\sin^2 \theta} \left( n \cos \theta E + L \right),
    \end{split}
    \label{eq:Theta}
\end{equation}
\begin{equation}
    R = R(r) =
    \left( E(a \chi + \Sigma) - aL \right)^2 - \Delta \left( K + r^2 + n^2 + (L-aE)^2 \right).
    \label{eq:R}
\end{equation}
Using (\ref{eq:4vel-r}), we can define the effective potential such that
\begin{equation}
    \frac{1}{2}(u^r)^2 + V_\mathrm{eff}(r,\theta) = \frac{1}{2}(E^2-1).
\end{equation}
As $r \rightarrow \infty$, the values $R(r)\rightarrow \frac{1}{2}(E^2-1)$ and $V_\mathrm{eff}(r,\theta) \rightarrow 0$. Notice that because of the square roots in (\ref{eq:R}) and (\ref{eq:Theta}), $R(r)\geq 0$ and $\Theta \geq 0$ must be satisfied. Consequently, for a test particle at $r \rightarrow \infty$, $E \geq 1$ must be true. With that, we call particles with $E < 1$ bounded particles, particles with $E = 1$ marginally bounded particles, and particles with $E > 1$ unbounded particles.

We can get the critical angular momentum of the particle using the "forward-in-time" condition $u^t > 0$ along the geodesic and plug that into equation (\ref{eq:4vel-t})
\begin{equation}
    L < E \frac{(r^2+n^2+a^2)^2\sin^2\theta - \chi^2\Delta }{a(r^2+n^2+a^2)\sin^2\theta - \chi\Delta}.
    \label{eq:angular-mom-forward-in-time}
\end{equation}
For $\Delta \rightarrow 0$ or $r \rightarrow r_{\pm}$, (\ref{eq:angular-mom-forward-in-time}) becomes
\begin{equation}
    L \leq \frac{E\left( r_\pm^2 + n^2 + a^2 \right)}{a}
\end{equation}
which gives an upper limit to the angular momentum of a particle near the horizon. This upper limit is the critical angular momentum $\hat{L}_\pm$, i.e. the maximum angular momentum at a given $E$ such that the 4-velocity is still normalized as the particle approaches the horizon. The $\pm$ subscript indicates which horizon it corresponds to ($r_\pm$). For extremal KNTN spacetime, $r_+ = r_- = M$ and consequently,  $\hat{L}_\pm = \hat{L}$.
\begin{equation}
    \begin{gathered}
        \Hat{L}_\pm = \frac{E\left( r_\pm^2 + n^2 + a^2 \right)}{a},\\
        \Hat{L} = \frac{E(2a^2 + Q^2)}{a}.
    \end{gathered}
    \label{eq:crit-angular-mom}
\end{equation}
In this spacetime, center-of-mass energy of two neutral test particles is \cite{Zakria_2015}
\begin{equation}
    \frac{E_{cm}}{\sqrt{2 m_1 m_2}} = \sqrt{ \frac{(m_1-m_2)^2}{2m_1m_2}
    + \frac{F(r,\theta) - G(r,\theta) - H(r,\theta)}{I(r,\theta)}}
    \label{eq:E_cm-KNTN}
\end{equation}
where
\begin{equation}
    \begin{split}
        F(r,\theta) = {} & \Delta \Sigma \sin^2\theta - (\Delta - a^2 \sin^2\theta) L_1 L_2 + \left( (\Sigma+a\chi)^2 \sin^2\theta - \chi^2 \Delta \right) E_1 E_2 \\
        & + \left( \chi \Delta - a(\Sigma-a\chi) \sin^2\theta \right) (L_1 E_2 + L_2 E_1),\\
        G(r,\theta) = {} & \sin^2\theta \sqrt{R_1(r) R_2(r)},\\
        R_i(r) = {} & \left( E_i(a \chi + \Sigma) - aL_i \right)^2 - \Delta \left( K_i + r^2 + n^2 + (L_i-aE_i)^2 \right),\\
        H(r,\theta) = {} & \Delta \sin^2\theta \sqrt{\Theta_1(\theta) \Theta_2(\theta)},\\
        \Theta_i(\theta) = {} & K_i - \cos^2\theta \left( (1-E_i^2)a^2 + \frac{L_i^2}{\sin^2\theta} \right) - 2an \cos \theta (1-2E_i^2) \\
        & - \frac{4an \cos \theta E_i}{\sin^2 \theta} \left( n \cos \theta E_i + L_i \right),\\
        I(r,\theta) = {} & \Delta \Sigma \sin^2 \theta,
    \end{split}
\end{equation}
with the subscript $i$ denotes the $i$-th particle. For a collision near the horizon $r \rightarrow r_\pm$, (\ref{eq:E_cm-KNTN}) becomes
\begin{align*}
    \left.\frac{E_{cm}}{\sqrt{2 m_1 m_2}}\right|_{r\rightarrow r_\pm} = {} &
    \left\{ \frac{(m_1-m_2)^2}{4m_1m_2}+1+\frac{1}{4(\hat{L}_{\pm1}-L_1)(\hat{L}_{\pm2}-L_2)} \left[\vphantom{\frac{r^2}{r^2}}[(\hat{L}_{\pm1}-L_1) - (\hat{L}_{\pm2}-L_2)]^2 \right.\right.\\
    & + \frac{1}{r_\pm^2 + (n+a\cos\theta)^2} \left( \frac{(r_\pm^2+n^2)^2}{(r_\pm^2+n^2+a^2)^2}(L_1 \hat{L}_{\pm2} - L_2 \hat{L}_{\pm1})^2 + K_2(\hat{L}_{\pm1 - L_1})^2 \right.\\
    & \left. \left. + K_1(\hat{L}_{\pm2}-L_2)^2 - a \cos\theta(2n+a\cos\theta) [(\hat{L}_{\pm1}-L_1)^2 + (\hat{L}_{\pm2}-L_2)^2] \vphantom{\frac{r^2}{r^2}} \right) \right]\\
    & - \frac{1}{2\left( r_\pm^2 + (n+a\cos\theta)^2 \right) \sin^2\theta} \left[ \cos^2\theta L_1 L_2 + \frac{2an\cos\theta (L_1 \hat{L}_{\pm2} + L_2 \hat{L}_{\pm1})}{r_\pm^2 + n^2 + a^2} \right.\\
    & \left. \left. + \frac{a^2 [(a\sin^2\theta - 2n\cos\theta)^2 - a^2 \sin^2\theta]}{(r_\pm^2 + n^2 +a^2)^2} \hat{L}_{\pm1} \hat{L}_{\pm2} + \sin^2\theta \sqrt{\Theta_1(\theta) \Theta_2(\theta)} \right] \right\}^{\frac{1}{2}}, \numberthis
    \label{eq:E_cm-KNTN-horizon-limit}
\end{align*}
with $\hat{L}_{\pm i}$ is the critical angular momentum of the $i$-th particle that corresponds to the outer/inner horizon $r_\pm$. Notice that $E_{cm}$ goes to infinity when one of the particles has an angular momentum equals to its critical angular momentum $L_i = \hat{L}_{\pm i}$.

\section{Constant $\theta$ geodesics}
\label{sec:const-theta}

Here we discuss the constant $\theta$ geodesics since we are interested in checking the equatorial BSW effect. We need to consider two constraints, $u^\theta = 0$ and $\frac{\mathrm{d}u^\theta}{\mathrm{d}\tau}=0$ so that the coordinate $\theta$ does not change along the geodesic, which also ensure the existence of equatorial geodesics. We apply those constraints to the derivative of the square of (\ref{eq:4vel-theta}) w.r.t proper time $\tau$. The square of (\ref{eq:4vel-theta}) is
\begin{equation}
\label{eq:Th^2}
\Theta = (\Sigma u^\theta)^2.
\end{equation}
Differentiating (\ref{eq:Th^2}) w.r.t proper time $\tau$ gives us
\begin{gather*}
\frac{\te{d}\Theta}{\te{d}\tau} = \frac{\te{d}}{\te{d}\tau} (\Sigma u^\theta)^2,\\
\frac{\te{d}\Theta}{\te{d}\theta} u^\theta = 2\Sigma u^\theta \left[ \frac{\te{d}\Sigma}{\te{d}\tau} u^\theta + \Sigma \frac{\te{d}u^\theta}{\te{d}\tau} \right]. \numberthis
\label{eq:dTheta-dtheta}
\end{gather*}
Canceling out $u^\theta$ from both sides\footnote[1]{This might seem problematic since $u^\theta=0$. However, we get the same result if we apply the constraints to the Hamilton's equation for $\dot{P}_\theta$ where no such canceling out is needed. See appendix \ref{app} for an alternative method of applying constant-$\theta$ constraints.} and applying the constraints, we conclude that 
\begin{equation}
	\label{eq:dTh-dth=0}
	\frac{\te{d}\Theta}{\te{d}\theta}=0
\end{equation}
must be satisfied.
Meanwhile, the constraint $u^\theta = 0$ applied to (\ref{eq:Th^2}) yields
\begin{equation}
	\label{eq:Th=0}
	\Theta = 0.
\end{equation}
The general equations of motion have 3 free parameters, namely $E,\ L,\ \te{and}\ K$. By applying the two constraints, $u^\theta = 0$ and $\frac{\te{d}u^\theta}{\te{d}\tau}=0$, the particle now only has 1 free parameter. We are going to assert the constraints onto $L\ \te{and}\ K$ and $E$ will be the only free parameter of the particle.

\section{BSW effect in KNTN spacetime}
\label{sec:BSW-KNTN}
\subsection{Equatorial plane}
\label{subsec:equatorial-BSW}
Now let us discuss the equatorial BSW effect in KNTN spacetime.
Using $\theta=\frac{\pi}{2}$, equation (\ref{eq:Th=0}) gives us
\begin{equation}
	\label{eq:K-constraint-equatorial}
	\tilde{K} = 0,
\end{equation}
where the \textit{tilde} denotes the constant-$\theta$ class geodesics. From (\ref{eq:dTh-dth=0}), while also applying $K=\tilde{K}=0$, we get
\begin{equation}
	\label{eq:L-constraint-equatorial}
	\tilde{L} = \frac{a(2E^2-1)}{2E},
\end{equation}
which is the same specific angular momentum found in \cite{Cebeci_2016,Cebeci_2019} for a particle that is confined to move over the equatorial plane. As mentioned above about the \textit{tilde} notation, $\tilde{K}$ and $\tilde{L}$ are the values of $K$ and $L$ respectively for a particle with constant $\theta$ geodesics.

Since $E_{cm}$ diverges only if $L_i = \hat{L}_{\pm i}$ and a particle that moves in the equatorial must have $L_i = \tilde{L}_i$, we check whether $\hat{L}_{\pm} - \tilde{L} = 0$ can be satisfied. However, we can see that
\begin{equation}
    \Hat{L}_\pm - \Tilde{L} = \frac{2E^2 (r_\pm^2+n^2) + a^2}{2aE}.
    \label{eq:diff-L}
\end{equation}
For $M,\ n,\ a,\ Q,\ E \in \mathbb{R}_{\not=0}$, the value $\hat{L}_\pm - \tilde{L} \not=0$ which means BSW effect is not possible if the particles are moving on the equatorial plane. The same conclusion can be made for extremal horizon with $\hat{L}_\pm = \hat{L}$ and $r_\pm = M$.

\begin{figure}
	\centering
	\includegraphics[width=.7\linewidth]{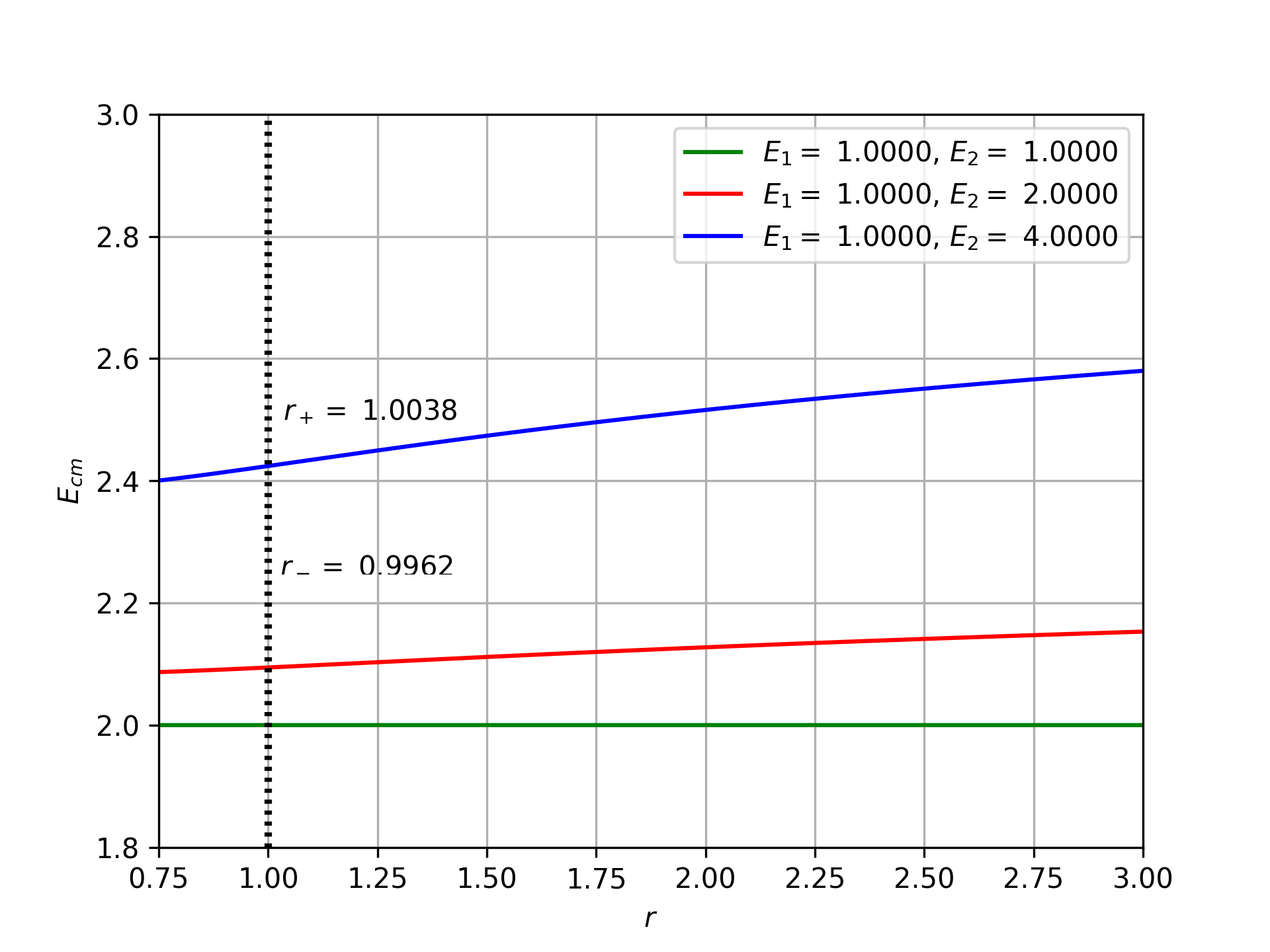}
	\caption{Plot of $E_{cm}(r)$ with $M=1,\ m_1=m_2=1,\ a=0.8,\ n=0.4,\ Q=0.7211$, $E_1=1,\ \te{and}\ E_2=1,2,4$}
	\label{fig:E_cm(r)-eq}
\end{figure}

Plugging in $L_1 = L_2 = \tilde{L}_\pm$, $K_1 = K_2 = \tilde{K} = 0$, and $\theta = \frac{\pi}{2}$ into (\ref{eq:E_cm-KNTN-horizon-limit}) yields
\begin{equation}
    \begin{aligned}
        \left.\frac{E_{cm}}{2\sqrt{m_1m_2}} \right|_{r\rightarrow r_\pm} = {} &\left[ \frac{(E_1-E_2)^2 \left[ 4(r_\pm^2+n^2)^2E_1^2E_2^2 + a^2 (r_\pm^2+n^2) (E_1 - E_2)^2 + a^4 \right] }{4E_1 E_2 \left[2(r_\pm^2+n^2)E_1^2+a^2\right]\left[2(r_\pm^2+n^2)E_2^2+a^2\right]} \right.\\
        & \left. + \frac{(m_1-m_2)^2}{4m_1m_2} + 1 \right]^\frac{1}{2}.
    \end{aligned}
\end{equation}
This leads to the same conclusion that $E_{cm}$ does not diverge for non-zero parameters $M,\ n,\ a,\ Q$, and $E$, i.e. contrary to \cite{Zakria_2015}, equatorial BSW effect is not possible in KNTN spacetime. As an example, figure \ref{fig:E_cm(r)-eq} is the plot of $E_{cm}$ as a function of $r$ with $M=1$, $m_1 = m_2 = 1$, $a=0.8$, $n=0.4$, $Q=0.7211$, $E_1=1$ and $E_2=1,2,4$. These spacetime parameters were used in \cite{Zakria_2015} to show the supposed equatorial BSW effect. However, as seen above, $E_{cm}$ does not diverge near the horizon.

\subsection{Other constant $\theta$ geodesics}
\label{subsec:any-const-theta}
Since we established that no BSW effect can happen for two neutral test particles that are moving only in the equatorial plane, we decided to take a look at other values of constant $\theta$ for $\theta \in [0,\pi]$. Solving equation (\ref{eq:dTh-dth=0}) for $L$ yields
\begin{equation}
    \Tilde{L} =
    \begin{dcases*}
        \text{Undefined}, &$\theta = 0$ or $\theta = \pi$,\\
        \frac{a(2E^2-1)}{2E}, & $\theta = \frac{\pi}{2}$,\\
        \frac{-nE(1+\cos^2\theta) - \sin^2\theta \sqrt{\left[(E^2-1)a \cos\theta+nE^2\right](n+a\cos\theta)}}{\cos\theta}, &otherwise.
    \end{dcases*}
    \label{eq:L-constraint-non-equatorial}
\end{equation}
The quantity $\tilde{L}$ is undefined because $\frac{\te{d}\Theta}{\te{d}\theta} \rightarrow \infty$ when $\theta = 0$ or $\theta=\pi$. Substituting (\ref{eq:L-constraint-non-equatorial}) into (\ref{eq:Th=0}) and solving for $K$ gives us
\begin{equation}
    \Tilde{K} = 
    \begin{dcases*}
        \text{Undefined}, &
        $\theta=0$ or $\theta=\pi$\\
        0, & $\theta = \frac{\pi}{2}$\\
        \left.
        \begin{aligned}
            &- 2nE \sin^2\theta \sqrt{\left[a(E^2-1)\cos\theta + nE^2\right](n+a\cos\theta)}\\
            & - a^2 (E^2-1) \cos^4\theta - \left[2an \left(E^2 - \frac{1}{2}\right) \cos^3\theta - 2 n^2 E^2\right]\\
            & \times \left( \cos^2\theta+1 \right)
        \end{aligned}\right\}
        & otherwise.
    \end{dcases*}
    \label{eq:K-constraint-non-equatorial}
\end{equation}
The quantity $\tilde{K}$ is also undefined because $\Theta \rightarrow \infty$ when $\theta = 0$ or $\theta=\pi$. $\tilde{L}$ and $\tilde{K}$ are the values of $K$ and $L$ respectively for a particle with constant $\theta$ geodesic.

Since $\tilde{K}$ and $\tilde{L}$ are undefined at $\theta = 0$ and $\theta=\pi$ and since BSW effect cannot happen at $\theta=\frac{\pi}{2}$, we are just going to look at the case where $0<|\theta - \frac{\pi}{2}|<\frac{\pi}{2}$. To analyze the possibility of BSW effect in this region, we define $\tilde{E}_{BSW_\pm}$ as the value of $E$ such that $\tilde{L} = \hat{L}_\pm$.
\begin{equation}
    \begin{aligned}
        \left(\Tilde{E}_{BSW_\pm}\right)^2 = {}& \left[ -a^3 \cos\theta (n+a\cos\theta) \sin^4\theta \right] \left\{ 2an \left(r_\pm^2+n^2+a^2\right) \cos^3\theta \right. \\
        & \left. + \left[a^4\left(1-\sin^4\theta\right)+2\left(r_\pm^2+3n^2\right)a^2 +\left(r_\pm^2+n^2\right)^2\right]\cos^2\theta \right. \\
        & \left. - 2an\left[a^2\sin^4\theta - \left(r_\pm^2+n^2+a^2\right)\right] \cos\theta  \right\}^{-1}.
    \end{aligned}
    \label{eq:E-constraint-BSW}
\end{equation}
We can see that $\tilde{E}_{BSW_\pm} \in \mathbb{R}$ only for certain values of $\theta$. Also the condition that $R(r) \geq 0$ must also be kept in mind. It turns out checking for the possibility of BSW effect in this case analytically is really troublesome, so we decided to approach this numerically.

We are going to do the numerical calculations as follows. We first set the spacetime parameters $M,\ n,\ a$, and $Q$ as well as the particles' masses $m_1=m_2=1$. Within $10^{-3}$ of precision in $\theta \in [0,\pi]$, we calculate the value of $\tilde{E}_{BSW_\pm}$. For every $\tilde{E}_{BSW_\pm} \in \mathbb{R}$, we set $E_2 = \tilde{E}_{BSW_\pm}$ which consequently sets $L_2 = \tilde{L} = \hat{L}_\pm$ and $K_2 = \tilde{K}|_{\tilde{E}_{BSW_\pm}}$. We choose the value of $E_1$ which then determines $L_1 = \tilde{L}|_{E_1}$ and $K_1 = \tilde{K}|_{E_1}$. Now that we have all of the parameters we need, we can plot $E_{cm}(r)$ using (\ref{eq:E_cm-KNTN}). For visualization purposes, we are going to evenly choose 5 or 6 values of $\theta$ from the range of $\theta$ that yields $\tilde{E}_{BSW_\pm} \in \mathbb{R}$. This will not give us the full picture but it might give us a glimpse on how $E_{cm}(r)$ behaves for a handful of parameters.
The followings are some numerical investigation.

\begin{figure}
	\centering
	\includegraphics[width=.7\linewidth]{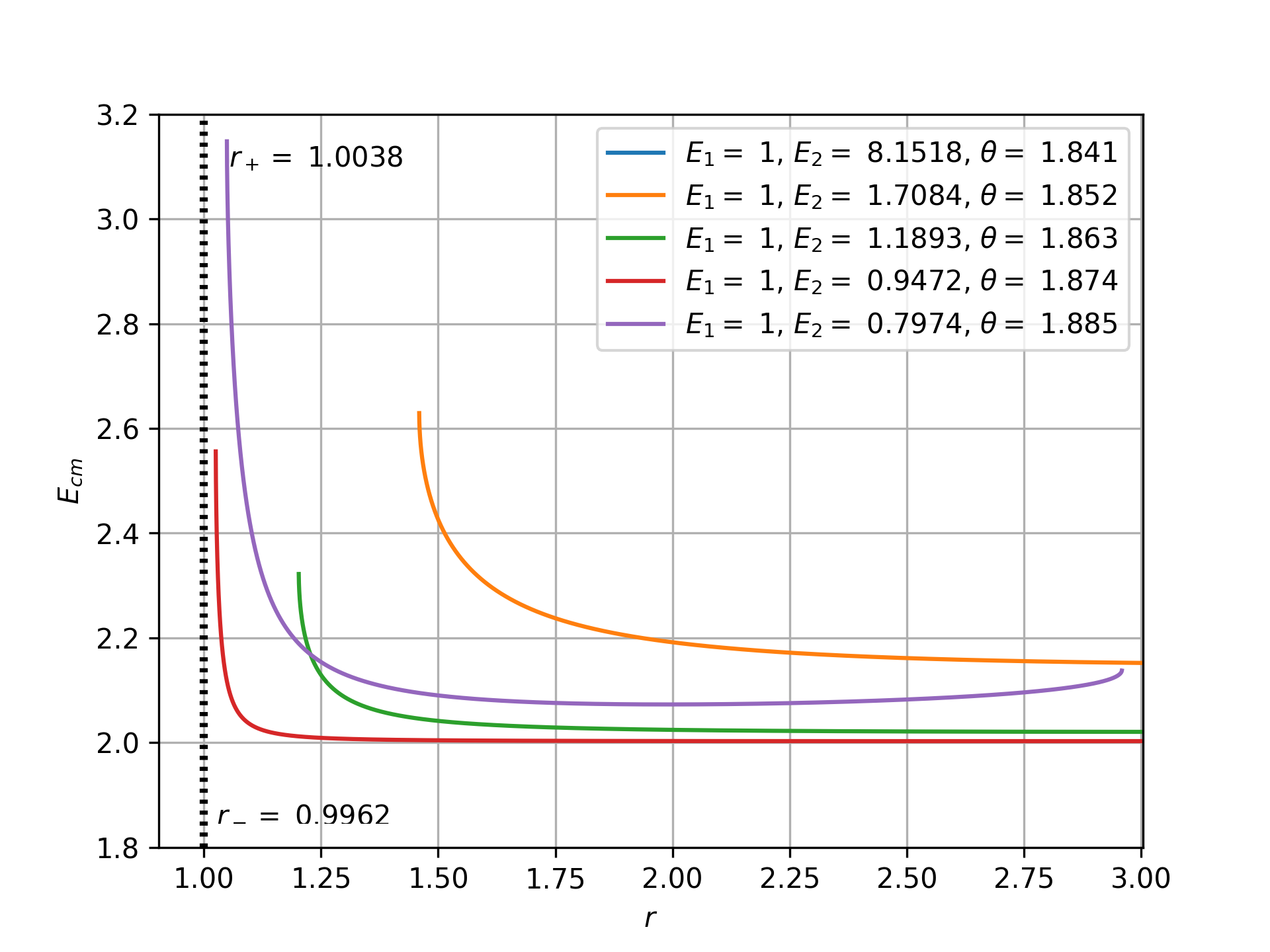}
	\caption{Plot of $E_{cm}(r)$ for a collision near the outer horizon $r_+$ of the non-extremal KNTN metric. The parameters used are $M=1,\ m_1=m_2=1,\ a=0.8,\ n=0.4,\ Q=0.7211,\ E_1=1$, and $E_2 = \tilde{E}_{BSW_+}$. $\tilde{E}_{BSW_+} \in \mathbb{R}$ if $\theta \in [1.841, 1.893]$.}
	\label{fig:E_cm(r)-noneq-outter1}
\end{figure}

\begin{figure}
	\centering
	\includegraphics[width=.7\linewidth]{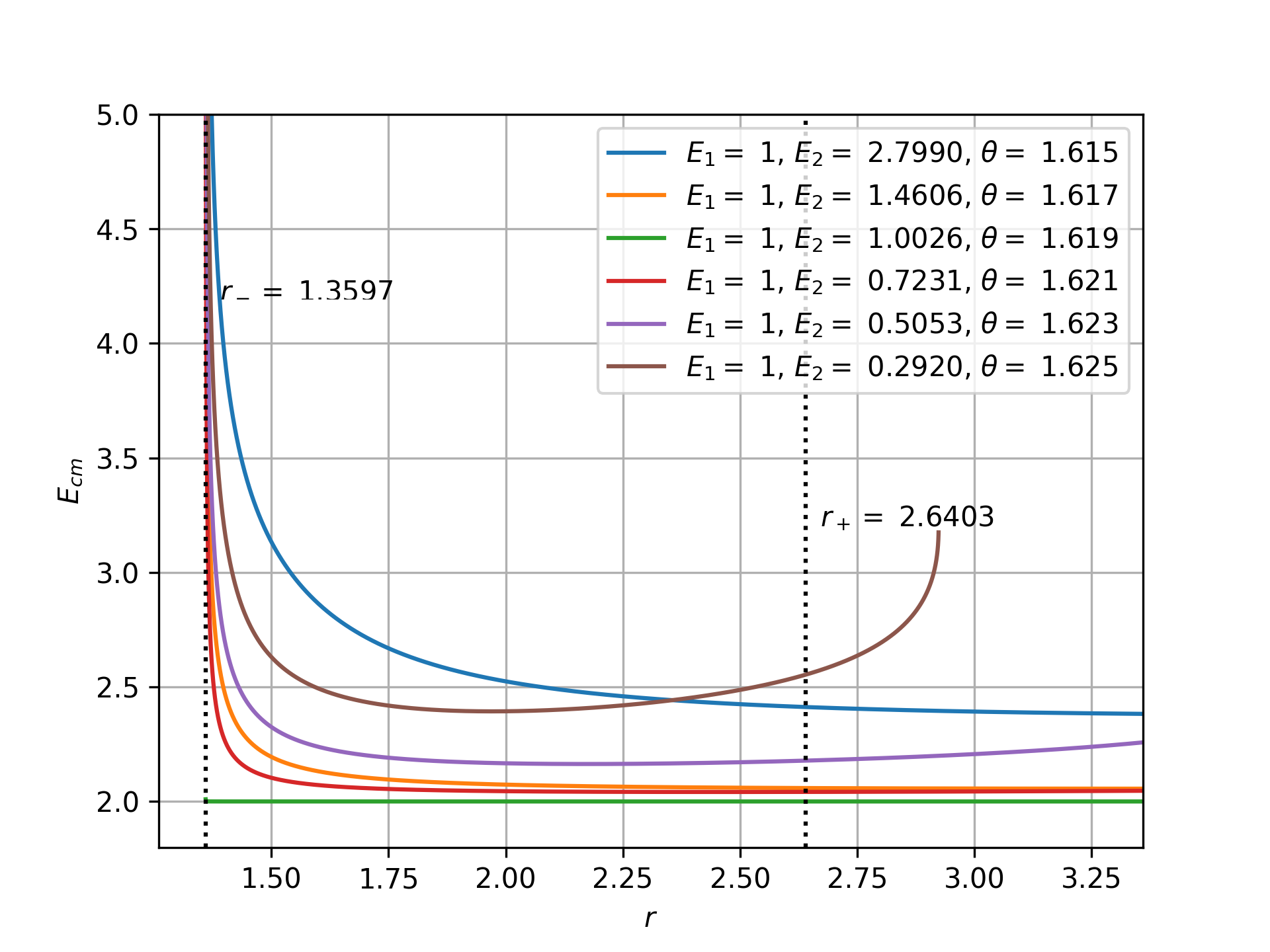}
	\caption{Plot of $E_{cm}(r)$ for a collision near the inner horizon $r_-$ of the non-extremal KNTN metric. The parameters used are $M=2,\ m_1=m_2=1,\ a=1.8,\ n=0.1,\ Q=0.6,\ E_1=1$, and $E_2 = \tilde{E}_{BSW_-}$. $\tilde{E}_{BSW_-} \in \mathbb{R}$ if $\theta \in [1.615, 1.626]$.}
	\label{fig:E_cm(r)-noneq-inner}
\end{figure}

\begin{figure}
	\centering
	\includegraphics[width=.7\linewidth]{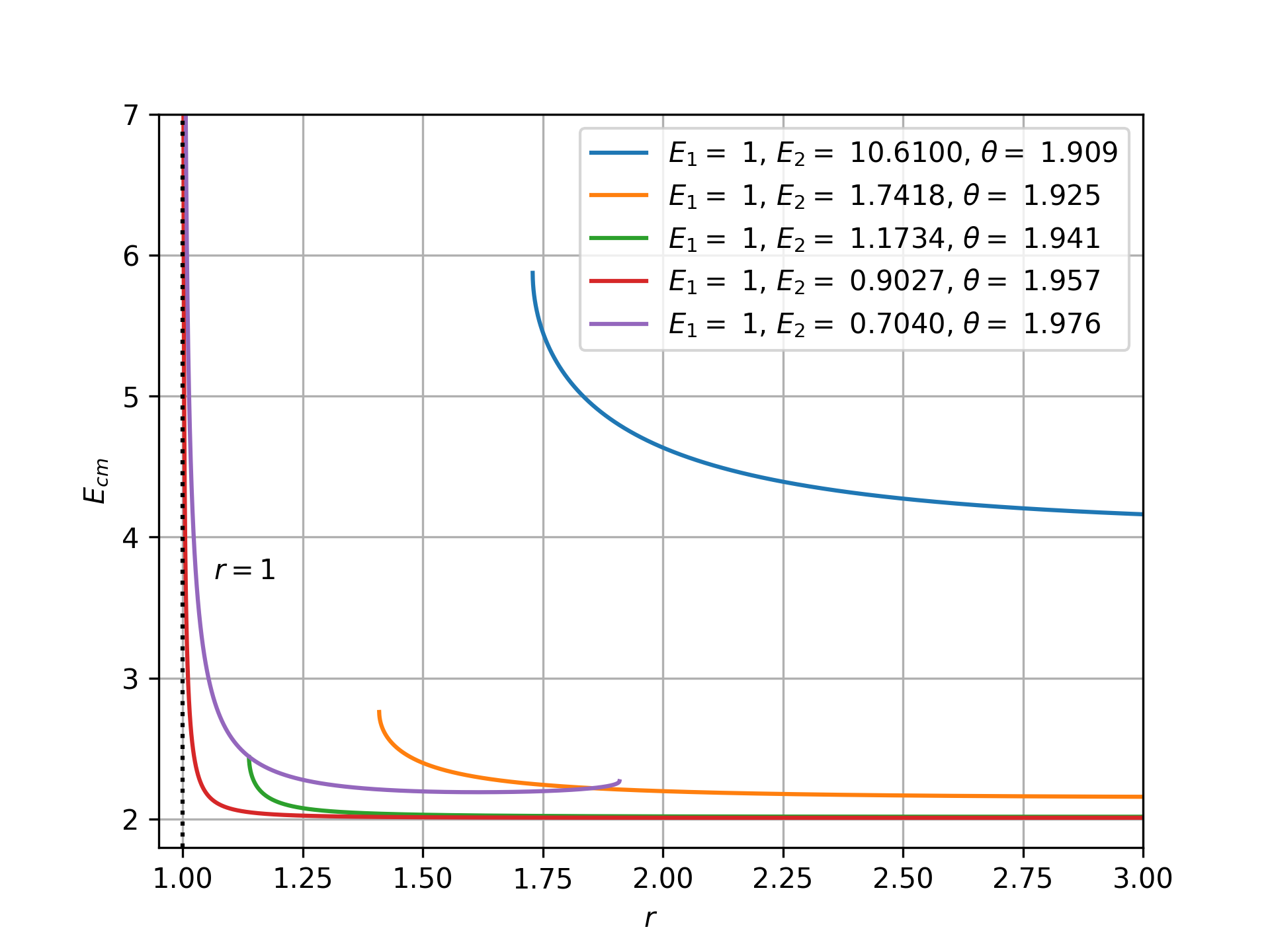}
	\caption{Plot of $E_{cm}(r)$ for a collision near the extremal horizon $r_\pm = M$ of the extremal KNTN metric. The parameters used are $M=1,\ m_1=m_2=1,\ a=1.06,\ n=\sqrt{0.2836},\ Q=0.4,\ E_1=1$, and $E_2 = \tilde{E}_{BSW_+}$. $\tilde{E}_{BSW_+} \in \mathbb{R}$ if $\theta \in [1.909, 1.991]$.}
	\label{fig:E_cm(r)-noneq-extremal}
\end{figure}

\begin{figure}
	\centering
	\includegraphics[width=.7\linewidth]{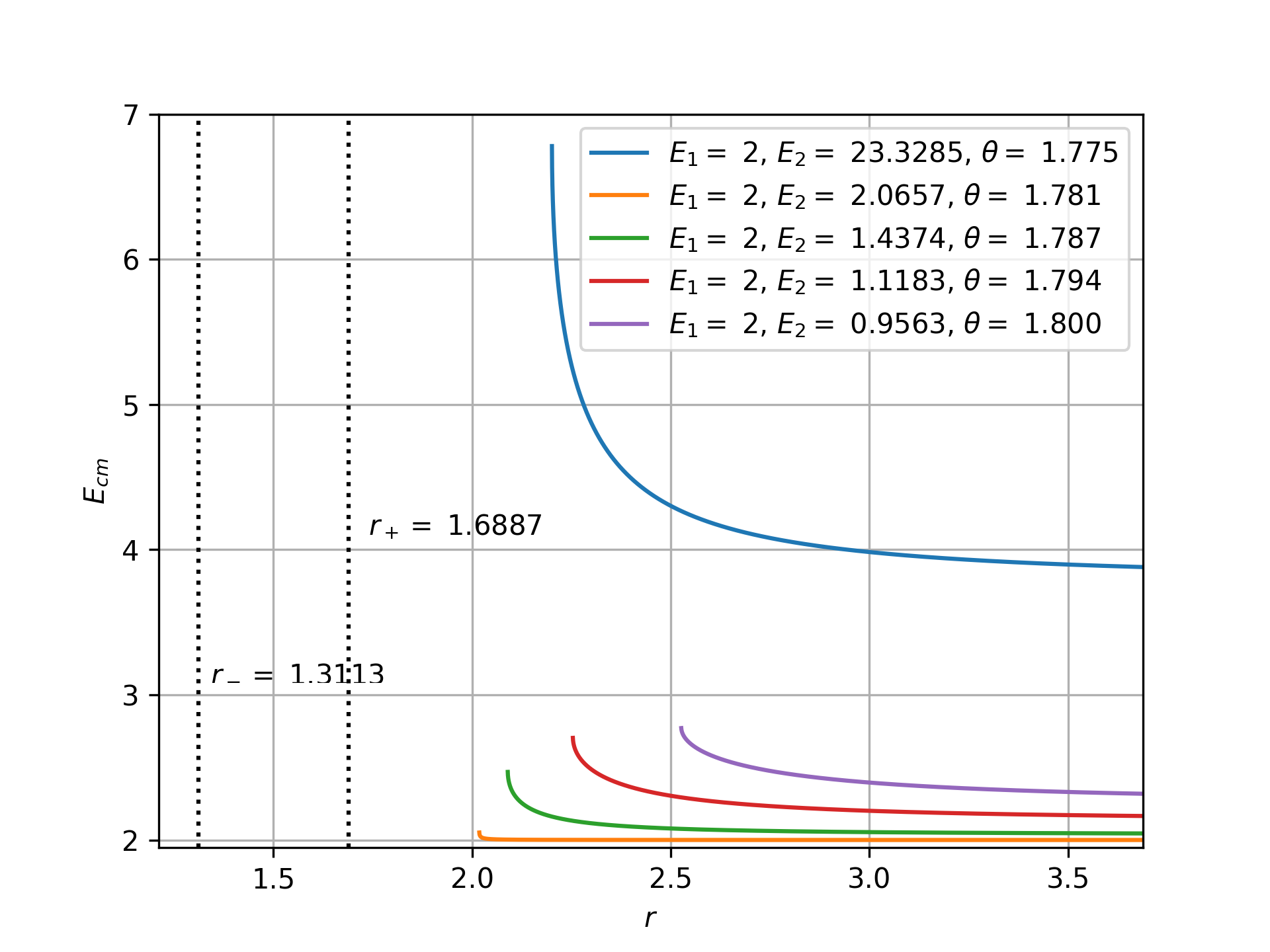}
	\caption{Plot of $E_{cm}(r)$ for a collision near the outer horizon $r_+$ of the non-extremal KNTN metric. The parameters used are $M=1.5,\ m_1=m_2=1,\ n=0.5,\ a=1.3,\ Q=0.88,\ E_1=1$, and $E_2 = \tilde{E}_{BSW_+}$. $\tilde{E}_{BSW_+} \in \mathbb{R}$ if $\theta \in [1.775, 1.804]$.}
	\label{fig:E_cm(r)-noneq-outter2}
\end{figure}

\begin{figure}
	\centering
	\includegraphics[width=.7\linewidth]{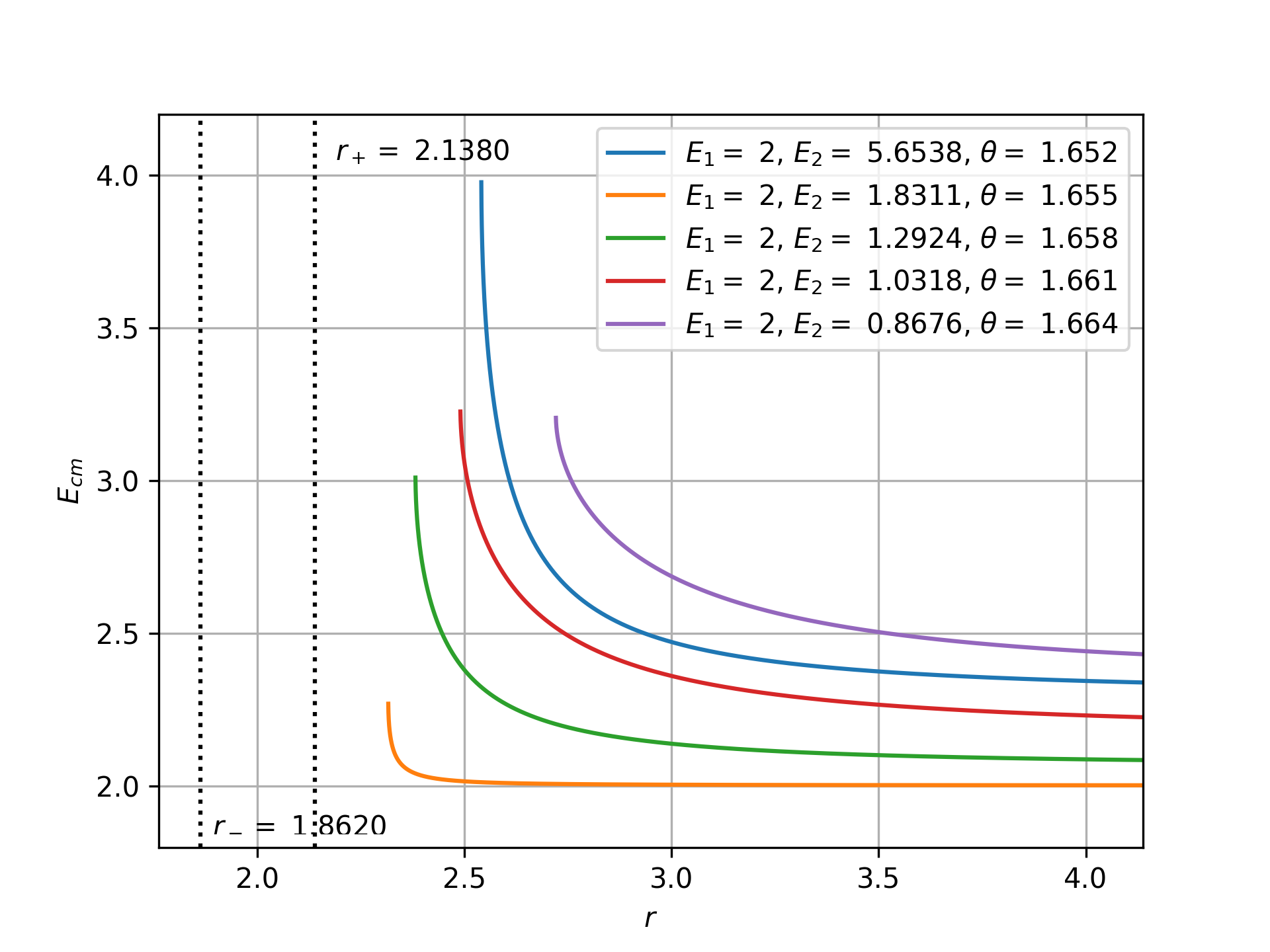}
	\caption{Plot of $E_{cm}(r)$ for a collision near the outer horizon $r_+$ of the non-extremal KNTN metric. The parameters used are $M=2,\ m_1=m_2=1,\ n=0.25,\ a=1.85,\ Q=0.788,\ E_1=1$, and $E_2 = \tilde{E}_{BSW_+}$. $\tilde{E}_{BSW_+} \in \mathbb{R}$ if $\theta \in [1.652, 1.666]$.}
	\label{fig:E_cm(r)-noneq-outter3}
\end{figure}

Figure \ref{fig:E_cm(r)-noneq-outter1} shows the plot of $E_{cm}$ as a function of $r$ for KNTN spacetime with parameters $M=1,\ m_1=m_2=1,\ a=0.8,\ n=0.4,\ Q=0.7211,\ E_1=1$, and $E_2 = \tilde{E}_{BSW_+}$. These parameters gives us $\tilde{E}_{BSW_+} \in \mathbb{R}$ if $1.841 \leq \theta \leq 1.893$. So $\theta$ is chosen evenly from that interval $[1.841, 1.893]$.
Figure \ref{fig:E_cm(r)-noneq-inner} shows the plot of $E_{cm}(r)$ for KNTN spacetime with parameters $M=2,\ m_1=m_2=1,\ a=1.8,\ n=0.1,\ Q=0.3,\ E_1=1$, and $E_2 = \tilde{E}_{BSW_-}$. These parameters gives us $\tilde{E}_{BSW_-} \in \mathbb{R}$ if $1.615 \leq \theta \leq 1.626$. So $\theta$ is chosen evenly from that interval $[1.615, 1.626]$.
Figure \ref{fig:E_cm(r)-noneq-extremal} shows the plot of $E_{cm}(r)$ for KNTN spacetime with parameters $M=1,\ m_1=m_2=1,\ a=1.06,\ n=\sqrt{0.2836},\ Q=0.4,\ E_1=1$, and $E_2 = \tilde{E}_{BSW}$. With these parameters, we get $\tilde{E}_{BSW} \in \mathbb{R}$ if $1.909 \leq \theta \leq 1.991$. Following the procedure, $\theta$ is chosen evenly from that interval $[1.909, 1.991]$.
So far, BSW effect is seen for a collision near the inner horizon $r_-$, which often thought as non-physical, in figure \ref{fig:E_cm(r)-noneq-inner} and near the extremal horizon in figure \ref{fig:E_cm(r)-noneq-extremal} when $E_2 < 1$.  BSW effect is yet to be found for a collision near the outer horizon of a non-extremal KNTN spacetime.
Because of that,
we tested a couple more non-extremal KNTN spacetime to probe a bit further for BSW effect near the outer horizon.
Parameters used in figure \ref{fig:E_cm(r)-noneq-outter2} are $M=1.5,\ m_1=m_2=1,\ n=0.5,\ a=1.3,\ Q=0.88,\ E_1=1,$ and $E_2 = \tilde{E}_{BSW_+}$ where $\tilde{E}_{BSW_+} \in \mathbb{R}$ if $\theta \in [1.775, 1.804]$.
Parameters used in figure \ref{fig:E_cm(r)-noneq-outter3} are $M=2,\ m_1=m_2=1,\ n=0.25,\ a=1.85,\ Q=0.788,\ E_1=1,$ and $E_2 = \tilde{E}_{BSW_+}$ where $\tilde{E}_{BSW_+} \in \mathbb{R}$ if $\theta \in [1.652, 1.666]$. 
In both results, we still cannot find BSW effect occurring. The numerical plots produced are as follows.


\newpage
\section{Conclusion}
In this paper, we revisit the BSW effect on equatorial plane in KNTN spacetime \cite{Zakria_2015}. In section \ref{subsec:equatorial-BSW}, contrary to the article \cite{Zakria_2015}, we concluded that no BSW effect can happen for particles that are strictly moving on the equatorial plane in KNTN spacetime. This differing result comes from a part of the constant-$\theta$ constraints that were overlooked, i.e. the $\frac{\te{d}u^\theta}{\te{d}\tau}=0$.
In section \ref{subsec:any-const-theta}, since the analytic studies are too complex, we switched to some numerical analysis to give us insights on the possibility of BSW effect. Admittedly, it is still inconclusive whether BSW effect can happen near the non-extremal outer horizon and near the extremal horizon with $E_2 \geq 1$ of a KNTN spacetime. 
However, it's notable that BSW effect is possible for a collision near the non-extremal inner horizon and for a collision near the extremal horizon if $\tilde{E}_{BSW} < 1$ although inner horizon is often thought to be non-physical. We doubt that BSW effect can take place in rotating spacetime with NUT parameter even on non-equatorial plane.
For future studies, it would be interesting to check for BSW effect in other rotating and charged spacetimes with NUT parameter such as Kerr-Sen-Taub-NUT spacetime \cite{Siahaan_2020_KSTN}, rotating and charged Taub-NUT-(A)dS spacetimes on a 3-brane \cite{Siahaan_2020_Brane}, and Kerr-Taub-NUT spacetime with Maxwell and dilaton fields \cite{Aliev_2008}.

\section*{Acknowledgment}
A massive thank you to Haryanto M. Siahaan for all of his guidance, patience, and knowledge.

\newpage
\appendix
\section{$\frac{\te{d}u^\theta}{\te{d}\tau}=0$ constraint using Hamilton's equation for $\dot{P}_\theta$}\label{app}

In section \ref{sec:const-theta}, we get that the $\frac{\te{d}u^\theta}{\te{d}\tau}=0$ constraint gives us equation (\ref{eq:dTh-dth=0}) which then gives (\ref{eq:L-constraint-non-equatorial}). We can get the same constraint for the angular momentum $L$ using Hamilton's equation for $\dot{P}_\theta$.
The Hamilton's equation we will use is
\begin{equation}
	\label{app-eq:hamiltons-pdot_theta}
	\frac{\te{d}P_\theta}{\te{d}\lambda} = \dot{P}_\theta = -\frac{\partial \mathcal{H}}{\partial \theta}
\end{equation}
where $\mathcal{H}$ is the hamiltonian (\ref{eq:hamiltonian}) which gives us
\begin{equation}
	\label{app-eq:PDotTheta}
	\dot{P}_\theta = \frac{1}{\Sigma^2 \Delta \sin^3(\theta)} \left[ \epsilon_0 + \epsilon_1 \cos(\theta) + \epsilon_2 \cos^2(\theta) + \epsilon_3 \cos^3(\theta) + \epsilon_4 \cos^4(\theta) + \epsilon_5 \cos^5(\theta) \right]
\end{equation}
where
\begin{align*}
	\epsilon_0 = {} & \left[ m^2 a^3 (E\chi-L)^2 + 2 E \Sigma m^2 a^2 (E\chi-L) + a \left\{ m^2 \left[ (\Sigma^2 - \Delta \chi^2)E^2 + 2 L E \Delta \chi - L^2 \Delta \right] \right. \right .\\
	& \left. \left. - \Delta P_\theta^2 - \Delta^2 P_r^2 \right\} - 2 E \Sigma \Delta m^2 (E \chi - L) \right], \numberthis \\
	\epsilon_1 = {} & a\ \epsilon_0, \numberthis \\
	\epsilon_2 = {} & -2n \left[ m^2 a^3 (E\chi-L)^2 + 2E \Sigma m^2 a^2 + a \left\{ \left[ \left( \Sigma^2 - \frac{\Delta}{2} \chi^2 \right) E^2 + L E \Delta \chi - \frac{\Delta}{2} L^2 \right] m^2 \right. \right. \\
	& \left. \left. \vphantom{\frac{\Delta}{2}} - \Delta P_\theta^2 - \Delta^2 P_r^2 \right\} - E \Sigma \Delta m^2 (E\chi-L) \right], \numberthis \\
	\epsilon_3 = {} & \frac{a}{n} \epsilon_2, \numberthis \\
	\epsilon_4 = {} & an \left[ a^2 m^2 (E\chi-L)^2 + 2E \Sigma m^2 a (E\chi-L) + E^2 \Sigma^2 m^2 - \Delta P_\theta^2 - \Delta^2 P_r^2 \right], \numberthis \\
	\epsilon_5 = {} & \frac{a}{n} \epsilon_4 \numberthis.
\end{align*}

With $P_r = \frac{m}{\Delta} \sqrt{R}$, $P_\theta = \frac{m}{\Sigma} \sqrt{\Theta}$, and (\ref{app-eq:PDotTheta}), solving $\dot{P}_\theta = 0$ for $L$ will give us (\ref{eq:L-constraint-non-equatorial}) from section \ref{subsec:any-const-theta}.


\begin{thebibliography}{10}
\newcommand{\enquote}[1]{``#1''}
\providecommand{\url}[1]{\texttt{#1}}
\providecommand{\urlprefix}{URL }
\expandafter\ifx\csname urlstyle\endcsname\relax
  \providecommand{\doi}[1]{doi:\discretionary{}{}{}#1}\else
  \providecommand{\doi}{doi:\discretionary{}{}{}\begingroup
  \urlstyle{rm}\Url}\fi
\providecommand{\eprint}[2][]{\url{#2}}

\bibitem{Griffiths_2009}
J.~B. Griffiths and J.~Podolsky,
\newblock {\em Exact Space-Times in Einstein{\textquotesingle}s General
  Relativity} (Cambridge University Press, 2009),
\newblock \doi{10.1017/cbo9780511635397}.

\bibitem{demianski1966combined}
M.~Demianski and E.~Newman,
\newblock Bull. Acad. Pol. Sci., Ser. Sci., Math., Astron., Phys. {\bf 14}, 653
  (1966).

\bibitem{Newman_1963}
E.~Newman, L.~Tamburino, and T.~Unti,
\newblock Journal of Mathematical Physics {\bf 4}, 915 (1963),
\newblock \doi{10.1063/1.1704018}.

\bibitem{Bini_2003}
D.~Bini, C.~Cherubini, R.~T. Jantzen, and B.~Mashhoon,
\newblock Class. Quantum Grav. {\bf 20}, 457 (2003),
\newblock \doi{10.1088/0264-9381/20/3/305}.

\bibitem{Al_Badawi_2006}
A.~Al-Badawi and M.~Halilsoy,
\newblock Gen Relativ Gravit {\bf 38}, 1729 (2006),
\newblock \doi{10.1007/s10714-006-0349-3}.

\bibitem{Ba_ados_2009}
M.~Ba{\~{n}}ados, J.~Silk, and S.~M. West,
\newblock Phys. Rev. Lett. {\bf 103}, 111102 (2009),
\newblock \doi{10.1103/physrevlett.103.111102}.

\bibitem{Berti_2009}
E.~Berti, V.~Cardoso, L.~Gualtieri, F.~Pretorius, and U.~Sperhake,
\newblock Phys. Rev. Lett. {\bf 103}, 239001 (2009),
\newblock \doi{10.1103/physrevlett.103.239001}.

\bibitem{Jacobson_2010}
T.~Jacobson and T.~P. Sotiriou,
\newblock Phys. Rev. Lett. {\bf 104}, 021101 (2010),
\newblock \doi{10.1103/physrevlett.104.021101}.

\bibitem{Galajinsky_2013}
A.~Galajinsky,
\newblock Phys. Rev. D {\bf 88}, 027505 (2013),
\newblock \doi{10.1103/physrevd.88.027505}.

\bibitem{Zaslavskii_2010}
O.~B. Zaslavskii,
\newblock Phys. Rev. D {\bf 82}, 083004 (2010),
\newblock \doi{10.1103/physrevd.82.083004}.

\bibitem{Harada_2011}
T.~Harada and M.~Kimura,
\newblock Phys. Rev. D {\bf 83}, 084041 (2011),
\newblock \doi{10.1103/physrevd.83.084041}.

\bibitem{Harada_2011_ISCO}
T.~Harada and M.~Kimura,
\newblock Phys. Rev. D {\bf 83}, 024002 (2011),
\newblock \doi{10.1103/physrevd.83.024002}.

\bibitem{Harada_2014}
T.~Harada and M.~Kimura,
\newblock Class. Quantum Grav. {\bf 31}, 243001 (2014),
\newblock \doi{10.1088/0264-9381/31/24/243001}.

\bibitem{Lake_2010}
K.~Lake,
\newblock Phys. Rev. Lett. {\bf 104}, 211102 (2010),
\newblock \doi{10.1103/physrevlett.104.211102}.

\bibitem{Grib_2011}
A.~Grib and Y.~Pavlov,
\newblock Astropart. Phys. {\bf 34}, 581 (2011),
\newblock \doi{10.1016/j.astropartphys.2010.12.005}.

\bibitem{Grib_2011_Near}
A.~A. Grib and Y.~V. Pavlov,
\newblock Gravit. Cosmol. {\bf 17}, 42 (2011),
\newblock \doi{10.1134/s0202289311010099}.

\bibitem{Grib_2010}
A.~A. Grib and Y.~V. Pavlov,
\newblock Jetp Lett. {\bf 92}, 125 (2010),
\newblock \doi{10.1134/s0021364010150014}.

\bibitem{Wei_2010}
S.-W. Wei, Y.-X. Liu, H.-T. Li, and F.-W. Chen,
\newblock J. High Energ. Phys. {\bf 2010}, 066 (2010),
\newblock \doi{10.1007/jhep12(2010)066}.

\bibitem{Liu_2013}
C.-Q. Liu,
\newblock Chinese Phys. Lett. {\bf 30}, 100401 (2013),
\newblock \doi{10.1088/0256-307x/30/10/100401}.

\bibitem{Chakraborty_2015}
C.~Chakraborty,
\newblock Eur. Phys. J. C {\bf 75}, 572 (2015),
\newblock \doi{10.1140/epjc/s10052-015-3785-y}.

\bibitem{Frolov_2012}
V.~P. Frolov,
\newblock Phys. Rev. D {\bf 85}, 024020 (2012),
\newblock \doi{10.1103/physrevd.85.024020}.

\bibitem{Liu_2011}
C.~Liu, S.~Chen, C.~Ding, and J.~Jing,
\newblock Phys. Lett. B {\bf 701}, 285 (2011),
\newblock \doi{10.1016/j.physletb.2011.05.070}.

\bibitem{Chakraborty_2014_Strong}
C.~Chakraborty and P.~Majumdar,
\newblock Class. Quantum Grav. {\bf 31}, 075006 (2014),
\newblock \doi{10.1088/0264-9381/31/7/075006}.

\bibitem{Zakria_2015}
A.~Zakria and M.~Jamil,
\newblock J. High Energ. Phys. {\bf 2015}, 147 (2015),
\newblock \doi{10.1007/jhep05(2015)147}.

\bibitem{Cebeci_2016}
H.~Cebeci, N.~Özdemir, and S.~{\c{S}}entorun,
\newblock Phys. Rev. D {\bf 93}, 104031 (2016),
\newblock \doi{10.1103/physrevd.93.104031}.

\bibitem{Cebeci_2019}
H.~Cebeci, N.~Özdemir, and S.~{\c{S}}entorun,
\newblock Gen Relativ Gravit {\bf 51}, 85 (2019),
\newblock \doi{10.1007/s10714-019-2569-3}.

\bibitem{Grenzebach_2014}
A.~Grenzebach, V.~Perlick, and C.~Lämmerzahl,
\newblock Phys. Rev. D {\bf 89}, 124004 (2014),
\newblock \doi{10.1103/physrevd.89.124004}.

\bibitem{Miller_1973}
J.~G. Miller,
\newblock J. Math. Phys. {\bf 14}, 486 (1973),
\newblock \doi{10.1063/1.1666343}.

\bibitem{Carter_1968a}
B.~Carter,
\newblock Phys. Rev. {\bf 174}, 1559 (1968),
\newblock \doi{10.1103/physrev.174.1559}.

\bibitem{Carter_1968b}
B.~Carter,
\newblock Commun.Math. Phys. {\bf 10}, 280 (1968),
\newblock \doi{10.1007/bf03399503}.

\bibitem{Siahaan_2020_KSTN}
H.~M. Siahaan,
\newblock Eur. Phys. J. C {\bf 80} (2020),
\newblock \doi{10.1140/epjc/s10052-020-08561-z}.

\bibitem{Siahaan_2020_Brane}
H.~M. Siahaan,
\newblock Phys. Rev. D {\bf 102}, 064022 (2020),
\newblock \doi{10.1103/physrevd.102.064022}.

\bibitem{Aliev_2008}
A.~N. Aliev, H.~Cebeci, and T.~Dereli,
\newblock Phys. Rev. D {\bf 77}, 124022 (2008),
\newblock \doi{10.1103/physrevd.77.124022}.

\end{thebibliography}

\providecommand{\href}[2]{#2}

\end{document}